\documentclass[pra,twocolumn,showpacs,superscriptaddress]{revtex4}
\usepackage[dvips]{graphicx}
\usepackage{amsmath,amsfonts,amssymb}
\usepackage{dcolumn}
\usepackage{bm}
\usepackage{epsfig}

\newcommand{\beq}{\begin{equation}}
\newcommand{\eeq}{\end{equation}}

\newcommand{\vecr}{{\bf r}}
\newcommand{\vecA}{{\bf A}}
\newcommand{\vecq}{{\bf q}}
\newcommand{\vecv}{{\bf v}}
\newcommand{\vecp}{{\bf p}}
\newcommand{\vecM}{{\bf M}}

\begin{document}

\title{Capture into Rydberg states and momentum distributions of ionized electrons}

\author{N.I.\ Shvetsov-Shilovski}
\affiliation{Moscow State Engineering Physics Institute, Kashirskoe Shosse 31, 115409, Moscow, Russia}
\author{S.P.\ Goreslavski}
\affiliation{Moscow State Engineering Physics Institute, Kashirskoe Shosse 31, 115409, Moscow, Russia}
\author{S.V.\ Popruzhenko}
\affiliation{Moscow State Engineering Physics Institute, Kashirskoe Shosse 31, 115409, Moscow, Russia}
\author{W.\ Becker}
\affiliation{Max-Born-Institut, Max-Born-Str.2a, 12489 Berlin, Germany}

\date{\today}

\begin{abstract}
The yield of neutral excited atoms and low-energy photoelectrons generated by the electron dynamics in the combined Coulomb and laser field after tunneling is investigated. We present results of Monte-Carlo simulations built on the two-step semiclassical model, as well as analytic estimates and scaling relations for the population trapping into the Rydberg states. It is shown that mainly those electrons are captured into bound states of the neutral atom  that due to their initial conditions (i) have moderate drift momentum imparted by the laser field and (ii) avoid strong interaction (``hard'' collision) with the ion. In addition, it is demonstrated that the channel of capture, when accounted for in semiclassical calculations, has a pronounced effect on the momentum distribution of electrons with small positive energy. For the parameters that we investigated its presence leads to a dip at zero momentum in the longitudinal momentum distribution of the ionized electrons.
\end{abstract}
%\keywords{Rydberg states, tunneling ionization, short laser pulses}
\maketitle

\section{Introduction}

The interaction of strong laser radiation with atoms and molecules generates a variety of highly nonlinear phenomena.  These include the extended photoelectron spectrum of above-threshold ionization (ATI), the excessive yield of doubly and multiply charged ions, the generation of very high harmonics of the driving field, and the yield of neutral atoms in excited Rydberg states (see \cite{Rev1,Rev2,Rev3,Fedorbook,DelKra} for reviews). The generally recognized transparent physical picture of the first three phenomena is based on the so-called direct ionization and rescattering scenario. According to it, electrons are promoted to the continuum via tunneling ionization and, while oscillating in the laser field, interact with their parent ions. Most of the ionized electrons, i.e. the ``direct'' electrons, are decelerated and deflected by the ion at small angles and finally contribute to the low-energy part of the photoelectron spectrum. In some rare cases, the ionized electron returns close to its parent ion and recombines emitting a high-frequency harmonic photon or kicks out one or several electrons, producing thereby a doubly or multiply charged ion. Yet another opportunity upon the close encounter is that the electron is elastically backscattered and thereafter acquires additional energy from the laser field. Such electrons contribute to the high-energy part of the photoelectron spectrum. The rescattering picture is explicitly realized in terms of complex ``quantum orbits'' or even simpler in terms of the two- and three-step semiclassical models. In the latter, real electron trajectories after ionization are calculated with Newton's equations.

The interpretation of the afore-mentioned processes is largely classical, based on trajectories obtained from Newton's equations. Population trapping in Rydberg states resulting in an excited neutral atom at the end of the laser pulse is an exception because until recently its  interpretation was purely quantum mechanical. The effect was considered as evidence of the existence of stabilization against ionization and studied theoretically and experimentally in many papers (see  \cite{Fedorbook,DelKra} and references therein). It is generally believed that there are two different mechanisms of stabilization: adiabatic stabilization (stabilization in the Kramers--Henneberger regime) and interference stabilization. The adiabatic approach is appropriate when the field frequency $\omega$ exceeds the atomic ionization potential $I$ so that the states in the discrete spectrum and the continuum are coupled by one-photon transitions. In experiments with optical lasers, such conditions were realized by initially  populating Rydberg states. The physical mechanism behind interference stabilization is associated with multiple Raman-type transitions between the Rydberg levels and the common continuum. Destructive interference of the transition amplitudes from these coherently populated states to the continuum suppresses ionization and, hence, serves to retain population in Rydberg states.

The recent paper \cite{Nubbemeyer} reports data on the yield of neutral excited He$^{\ast}$ atoms and singly charged ions He$^{+}$ from a gaseous target irradiated by 30-fs Titanium-Sapphire (Ti:Sa) laser pulses with intensities up to $10^{15}$ W/cm$^{2}$. The data, corrected for the radiative decay of neutral excited atoms on the way to the detector, fairly well agree with the results of extensive quantum-mechanical calculations. The latter were accomplished in two ways: by approximately solving the two-electron TDSE for helium exposed to a laser field and in single-active-electron (SAE) approximation. The excited neutral atom yield is roughly 20\% of the ion yield at lower intensities decreasing to about 10\% at higher intensities.

In addition, a two-step semiclassical model was employed for the first time for providing a simple picture of the dynamics resulting in the population of Rydberg states. Monte-Carlo calculations were performed with classical trajectories in the combined Coulomb potential and electric laser field. Those electrons are captured into bound states of the neutral atom that have negative total (kinetic plus Coulomb) energy at the end of the laser pulse. The energy distribution in the subset of bound trajectories was converted into the distribution in an effective principal quantum number $\left(E=-1/2n^{2}\right)$ and the latter was found to be in qualitative and even quantitative agreement with the one deduced from quantum calculations.

The implication of this study is that in Monte-Carlo simulations with classical trajectories in the laser field and the ionic Coulomb field not all electrons that tunneled out at some time are actually ionized after the end of the laser pulse. Some may be captured back into bound states. This outcome of the electron motion in the continuum was dubbed frustrated tunneling ionization (FTI) in \cite{Nubbemeyer}. Transient trapping in Rydberg states during the laser pulse was mentioned earlier in \cite{Yudin}. However, to the best of our knowledge, there are no comments on the formation of excited neutral atoms at the end of the pulse in the papers  where Monte-Carlo simulation of classical trajectories after tunneling was used to model the effect of Coulomb focusing on the rescattering processes \cite{focusing,Yudin,Comotis} or to investigate the low-energy photoelectron spectra \cite{Chen02,Dimitriou,Arbo}.

In this paper, we further exploit the semiclassical model: (i) to get deeper insight into the mechanism responsible for the yield of neutral exited atoms; (ii) to find out the scaling of this yield with the laser--atom parameters; and (iii) to understand the impact of the population trapping on the momentum distributions of the ionized electrons.

\section{Model}
In this section we will sketch our simulation technique with special emphasis on the details that are essential for the following.

Let us consider a neutral atom with the ionization potential $I$, irradiated by a short laser pulse with the duration $\tau_{L}=(2\pi/\omega)n_{p}$ and sine-square envelope, linearly polarized along the $x$ axis:
\begin{equation}
F(t)=F_{0}\sin^{2}(t/\tau_{L})\cos\left(\omega t\right),
\label{field}
\end{equation}
where $n_{p}$ is the number of cycles and $\omega$ is the carrier frequency.
At time $t_{0}$ the bound electron tunnels out of the atom quasistatically with the rate
\begin{multline}
    w\left(t_{0},v_{0}\right)\sim\left(\frac{F_{a}}{F\left(t_{0}\right)}\right)^{2}
    \frac{1}{\sqrt{1+v^{2}_{0}/2I}}\times\\
  \times\exp\left\{-\frac{2F_{a}}{3F\left(t_{0}\right)}\left[1+\frac{v^{2}_{0}}{2I}\right]^{3/2}\right\}.
    \label{rate}
\end{multline}
Here $I$ and $F_a=(2I)^{3/2}$ are the ionization potential and the characteristic atomic force of the bound state ($I=1/2$, $F_a=1$ for the ground state of hydrogen).
The rate (\ref{rate}) is obtained from the standard strong-field ionization amplitude \cite{Keldysh,FaisalReiss} calculated by the saddle-point method in the tunneling regime, when the Keldysh parameter $\gamma=\omega\sqrt{2I}/F\ll1$. Details of the calculations are presented elsewhere (see \cite{OurPRL,OurPRA}).
Atomic units $e=m=\hbar=1$ are used.

At the tunnel exit $x_{0}=I/F\left(t_{0}\right)$ the electron has zero longitudinal and nonzero initial transverse velocity $v_{0}$. The released electron moves along a classical trajectory under the action of the laser field and the Coulomb force of its residual ion. The initial conditions unambiguously determine its position $\vecr\left(t\right)=\vecr\left(t_{0},v_{0},t\right)$ and velocity $\vecq\left(t\right)=\vecq\left(t_{0},v_{0},t\right)$ at any time instant $t$. In the limit of large $t$, one obtains the asymptotic momentum $\vecp=\vecq\left(t_{0},v_{0},t\rightarrow\infty\right)$, which is identical with the  momentum measured by the detector. During the whole laser pulse $0\le t_{0} \le \tau_{L}$, an ensemble of trajectories is launched corresponding to different pairs $\left(t_{0},v_{0}\right)$ weighted with the probability (\ref{rate}).
No interference effects are accounted for.
In our simulations we employed about $1.5\cdot10^{6}$ trajectories regardless of the intensity. By solving the respective Newton equations one obtains $\vecr\left(t\right)$ and $\vecq\left(t\right)$ for each trajectory. The ratio of the number of trajectories in a particular bin of momentum space to the total number of trajectories is the momentum distribution at given time $t$ normalized to unity. It evolves in time and in the limit $t\rightarrow\infty$ can be compared with the experimental data.

%During the laser pulse, $0\le t \le \tau_{L}$, while the electron is moving in the two fields, its trajectory can be evaluated only numerically.
The picture described in the previous paragraph is not complete, however.
Indeed, if the Coulomb field of the nucleus is included into the Newton equations then, for some initial conditions $(t_0,v_0)$, the electron's momentum simply does not converge to any constant value but keeps oscillating in the limit $t\to\infty$.
Such trajectories correspond to those electrons whose total energy after the end of the laser pulse is negative
\begin{equation}
    E\left(\tau_{L}\right)=\frac{\vecq^{2}}{2}-\frac{1}{r}<0,
    \label{condition}
\end{equation}
so that they never reach a detector.
Following \cite{Nubbemeyer} we interpret these electrons as trapped in Rydberg states (captured electrons). Here $\vecq=\vecq(v_0,t_0,\tau_L)$ and
$\vecr=\vecr(v_0,t_0,\tau_L)$ are the electron velocity and its distance from the ion at the end of the laser pulse, respectively. Starting with the time $\tau_L$ the electron moves only in the Coulomb field, and its energy as well as the angular momentum are conserved. Thus, since this time the whole ensemble of trajectories can be subdivided into two subsets. The subset of trajectories satisfying condition (\ref{condition}) specifies the distribution of the bound electrons over the principal and the angular quantum numbers $n$ and $l$. The subset of trajectories with positive total energy determines the distribution of ionized electrons.

\section{Results and discussion}

\begin{figure}[t]
\begin{center}
\includegraphics[width=0.5\textwidth]{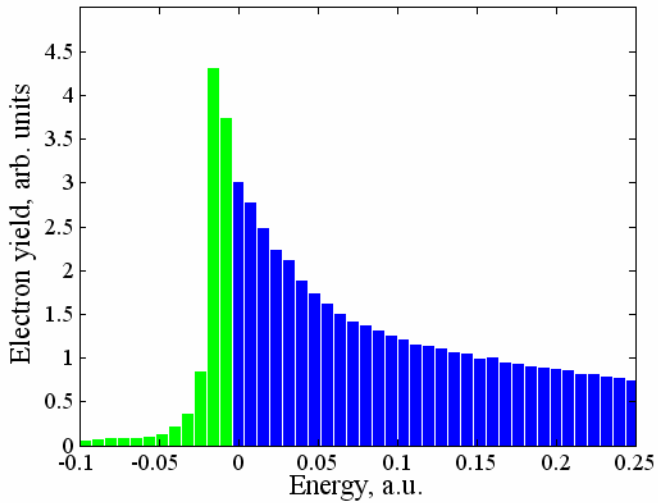}
\end{center}
\caption{The distribution of the total energy $E=\vecq^{2}/2-1/r$  of electrons tunneling out of an H-atom $\left(I=13.6 eV\right)$ at the end of a laser pulse with a duration of $n_{p}=5$ cycles, frequency $\omega=0.05$ a.u. and an intensity of $3.5\cdot10^{14}$W/cm$^{2}$. The number of electrons with negative energy (shown by green bars) is estimated at about  $15\%$, which is consistent with \cite{Nubbemeyer}.}
\label{enrhist}
\end{figure}

\begin{figure}[b]
\begin{center}
\includegraphics[width=0.5\textwidth]{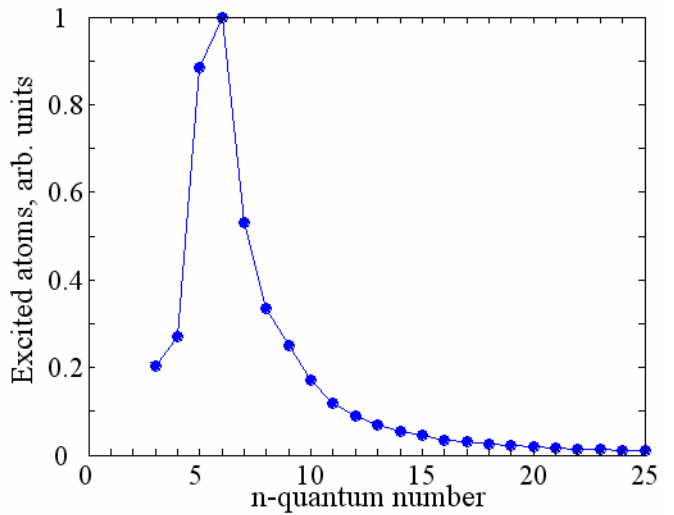}
\end{center}
\caption{The distribution of the captured electrons over the principal quantum number $n$. The parameters are those of Fig. 1.}
\label{distrn}
\end{figure}

\begin{figure}[t]
\begin{center}
\includegraphics[width=0.5\textwidth]{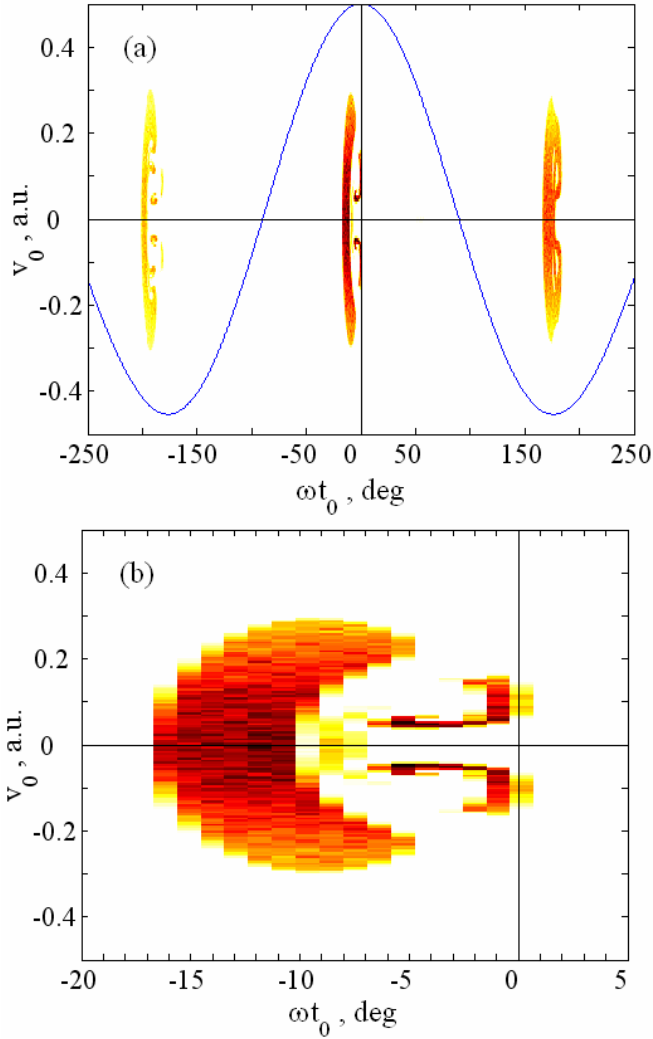}
\end{center}
\caption{(a) The 2D distribution of the captured electrons over the time of tunneling $t_{0}$ and initial  velocity $v_{0}$ for the parameters of Fig. 1. The laser field (\ref{field}) is shown by a blue line. (b) The same as in (a), but for the dominant maximum of the pulse. Dark means high electron yield.}
\label{cockroach}
\end{figure}

The results of our simulations are presented in Figs. \ref{enrhist}, \ref{distrn}, and \ref{cockroach}.

The energy distribution of the electrons at the time $\tau_{L}$ is shown in Fig. \ref{enrhist}. It can be seen from the figure that  the number of captured electrons corresponds to approximately $15\%$ of the number of ions, which agrees with the results obtained in \cite{Nubbemeyer}. The distribution of the trapped electrons over the principal quantum number is depicted in Fig.~\ref{distrn}. This distribution is similar to those evaluated in \cite{Nubbemeyer} by the solution of the time-dependent Schr\"odinger equation and by Monte-Carlo simulations: the maximum of the distribution corresponds to $n=6$.

Finally, Fig. \ref{cockroach} exhibits the areas of the plane $\left(t_{0},v_{0}\right)$ of initial conditions that lead to capture of the electron into Rydberg states. The profile of the laser pulse is also shown in the figure. Figure \ref{cockroach} reveals two main features. First, the capture does not take place for any initial velocity. It occurs only when $\left|v_{0}\right|$ is less than some maximum value (approximately 0.3 for the parameters of Fig. \ref{cockroach}). But this is not yet a sufficient condition.
Such a condition will be formulated in the next section, in which Fig.~\ref{cockroach} is explained and all necessary estimates are done.

\subsection{Mechanism and scalings of population trapping}

\begin{figure}[t]
\begin{center}
\includegraphics[width=0.35\textwidth,angle=-90.0]{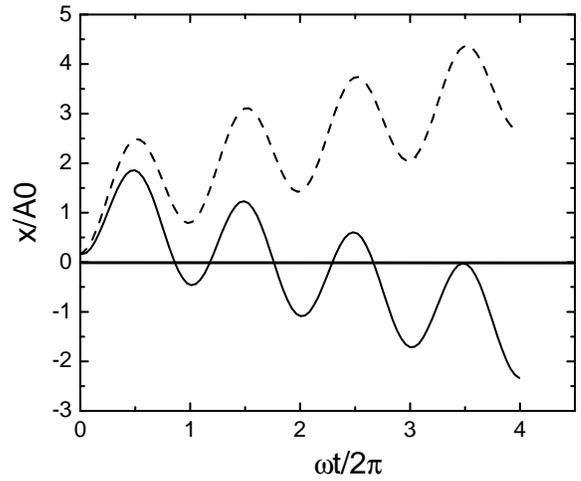}
\end{center}
\caption{Two electron trajectories in the field $F(t)=F_0\cos(\omega t)$ corresponding to two instants of ionization before and after the field maximum: $\omega t_0=-0.1$ (dashed curve) and $\omega t_0=0.1$ (solid curve) for the parameters of Fig.\ref{enrhist}. The coordinate is scaled by the amplitude $A_0=F_0/\omega^2$ of the electron oscillation in the laser field.}
\label{tr}
\end{figure}

To understand the results of Fig. \ref{cockroach} one should keep in mind that the momentum (velocity) $\vecq$ and the distance $\vecr$ in the capture condition (\ref{condition}) are actually correlated.
The larger the momentum, the larger is the electron distance and the smaller is the potential energy and, hence, the electron cannot be trapped.
With the equality sign, Eq.~(\ref{condition}) sets an upper limit for $q$.
Regarding smaller values for $q$, the electron can have a small drift momentum at the
end of the laser pulse only under certain conditions. 
The trajectories of electrons ionized near the field extrema with initially small drift
momentum along the laser field and with small transverse velocity are strongly affected by the attraction of the Coulomb field.
While such an electron oscillates in the laser field,
a single ``hard'' collision or a sequence of ``soft'' collisions
\cite{Yudin} are likely to increase its drift momentum so that, with
very high probability, it is not a candidate for capture anymore. 
On closer inspection, this holds for electrons that start their orbit \emph{after}
an extremum of the field. 
In this case, the drift momentum imparted by the laser field is directed towards the ion.
The net effect of the laser field is that these electrons are driven back to the ion at least once if not many times.
An example of such a trajectory is shown in Fig. \ref{tr} by the solid line.
The Coulomb force still further increases the drift momentum in the direction
towards the ion.
Inevitably electrons with large drift momentum return back to the ion and scatter depending on the magnitude of their impact parameter. 
The opportunity to avoid strong interaction with the ion and have sufficiently small drift momentum at the end of the laser pulse exists for electrons which start in the
continuum \emph{before} the field extremum.
Without the Coulomb field, while oscillating in the laser field they drift away from their
parent ion and never return to its vicinity (the trajectory shown by the dashed line in Fig. \ref{tr}). The Coulomb attraction reduces the drift momentum without reversing its direction, so that the trajectory remains ``on the average outgoing''.
This classification of trajectories is illustrated by Fig. \ref{tr}.

To support this general conclusion, let us consider the contribution to the capture process from a single half cycle of the laser field, for example, from the central maximum of the field (\ref{field}). With time counted from this maximum and for phases within the interval $-\pi /2<\omega t <\pi/2$, the field and its vector potential are simplified as $F\left(t\right)=F\cos \left( \omega t\right)$ and $A\left(t\right)=-\left(F/ \omega\right)\sin\left(\omega t\right)$.
The momentum at the end of the laser pulse of an electron tunneling at time $t_{0}$ with nonzero transverse velocity can be written as
\begin{equation}
  \vecq\equiv\vecq(t_0,\vecv_0,\tau_L) =\vecA\left(t_0\right)+\vecv_0+\vecp_C(t_0,\vecv_0).
  \label{Momentum}
\end{equation}
At $t>\tau_{L}$ the momentum (\ref{Momentum}) evolves in the Coulomb field. The trajectory starts at the tunnel exit $x_0=I/F\left(t_0\right)$ and the field $F\left(t_0\right)$ accelerates the electron away from it. For the laser--atom parameters of interest, the Coulomb force is small in comparison to the laser field already at the tunnel exit and rapidly decreases when the electron departs from the ion. For example, with the parameters of Figs. 1 and \ref{tr} and not too far from the field maximum one has $x_0\approx 5$ a.u., while the oscillation amplitude $A_0=F_0/\omega^2$ is about 40 a.u. This allows us to evaluate the contribution of the Coulomb force in Eq.(\ref{Momentum})

\begin{equation}
  \vecp_{C}\left(t_0,\vecv_0\right)=-\int^{+\infty}_{t_{0}}dt
  \frac{\vecr_{L}\left(t\right)}{r_L^{3}\left(t\right)}
  \label{momintegr}
\end{equation}
as the integral along the  initial part of the electron trajectory 
    \[
    \vecr_{L}\left(t\right)=\left\{x_{0}+\frac{1}{2}F\left(t_{0}\right)\left(t-t_{0}\right)^{2},
     v_0\left(t-t_{0}\right)\right\},
\]
and to extend the upper integration limit to infinity \cite{OurPRL}. Evaluation of the integral (\ref{momintegr}) for not very large transverse velocities, $v_0<\sqrt{2I}$, gives the drift momentum (\ref{Momentum}) in the plane defined by the directions of the field ($x$ axis) and of the transverse initial velocity ($\bot$ axis):
\begin{eqnarray}
  q_x&=&-\frac{F}{\omega}\sin\left(\omega t_{0}\right)-
  \pi\frac{F\cos\left(\omega t_{0}\right)}{\left(2I\right)^{3/2}},
  \label{qx}\\
q_{\bot}&=&v_0-2v_0\frac{\left|F\left(t_{0}\right)\right|}{\left(2I\right)^{2}}.
  \label{qy}
\end{eqnarray}
Except for the numerical factor, the Coulomb contribution to the drift momentum (\ref{qx}) is just the product of the Coulomb force at the tunnel exit, $-1 /x^{2}_{0}$, and the time interval
$\Delta t=\sqrt{2x_{0}/F\left(t_{0}\right)}$ over which the electron--ion distance doubles. 
Not surprisingly, this time interval coincides with the time of flight under the potential barrier, $x_0/\sqrt{2I}$, introduced by Keldysh.

For the next half period, the laser field changes its direction and pulls the outgoing electron back to the ion.
But, for $q_x>0$, the minimum  distance  in the $x$ direction between the electron and the ion will be larger than the tunnel exit $x_{0}$  by the drift displacement $q_x/\omega$. So, if this displacement is of the order of or larger than $x_{0}$, the Coulomb force upon the closest approach is less than the one at the time of tunneling, and its effect can be neglected. In such cases, the drift momentum given by (\ref{qx}) and (\ref{qy}) remains practically constant during the rest of the laser pulse. This allows us to estimate the electron--ion distance at the end of the laser pulse in Eq. (\ref{condition}) as

\begin{equation}
    r=q\left(\tau_L-t_0\right),
    \label{dist}
\end{equation}
where $q=\sqrt{q_x^2+ q_{\bot}^2}$. With the momentum and the distance at the end of the laser pulse known as functions of $t_{0}$  and $v_{0}$, the condition (\ref{condition}) determines the subspace of the initial parameters that result in capture to the bound states:

\begin{equation}
    q<\left(\frac{2}{\tau_{L}-t_{0}}\right)^{1/3}.
    \label{conditionest}
\end{equation}
It is immediately obvious that the number of trapped electrons decreases with increasing time that the electron spends in the laser field after ionization. This explains the difference in the number of electrons
captured from identical laser half-periods before and after  the field maximum, which is seen in Fig. \ref{cockroach}a.

Next, we try to obtain a rough analytical estimate for the relative yield of neutral excited atoms with respect to singly charged ions, i.e. the ratio $N^{\ast}/N^{+}$, in a short laser pulse. The numbers $N^{\ast}$ and $N^{+}$ can be evaluated by integrating the rate (\ref{rate}) over the respective parts of the $\left(t_{0},v_{0}\right)$-plane shown in Fig. \ref{cockroach}a. To a fair approximation, the integration can be performed over the initial conditions related to the central maximum of the field. Furthermore, applying the theorem of the mean to both integrals, we assume that the rates (\ref{rate}) do not differ essentially and cancel out in their ratio, which thereby is estimated as $N^{\ast}/N^{+}\approx\Sigma^{\ast}/\Sigma^{+}$. Here $\Sigma^{\ast}$ and $\Sigma^{+}$ are the areas of the $\left(t_{0},v_{0}\right)$ plane effective for population trapping and real ionization, respectively.  For ions from the central maximum of the field we have $\Sigma^{+}\approx \pi\times\sqrt{2I}\sqrt{F/F_{a}}$, where the second factor is the width of the rate (\ref{rate}) in the initial velocity.

For the neutrals, $\Sigma^{\ast}$ is just the area of the crescent-shaped region in Fig. \ref{cockroach}b. First, we find the interval of start times that are favorable for capture for $v_{0}=0$. Its left end $t_{0<}$ is found by equating the two sides of Eq. (\ref{conditionest}) under the assumption that we deal with an ``on the average outgoing trajectory'', i.e. $q_x>0$. The resulting equation is
\begin{equation}
  -\sin\left(\omega t_{0}\right)-\lambda \cos\left(\omega t_{0}\right)=\frac{\omega}{F}\left(\frac{2}{\tau_{L}-t_{0}}\right)^{1/3}
  \label{eqt0<}
\end{equation}
where $\lambda=\pi\omega/\left(2I\right)^{3/2}$. In the middle of the laser pulse, $\tau_{L}-t_{0}\approx \tau_{L}/2$, and, typically, the r.h.s. of Eq. (\ref{eqt0<}) is small in comparison with unity.  To first order in $\omega t_{0}<1$, the solution within the interval $-\pi/2<\omega t_{0}< \pi/2$ is

\begin{equation}
  \omega t_{0}\equiv \omega t_{0<}\approx-\lambda-\frac{\omega}{F}\left(\frac{4}{\tau_{L}}\right)^{1/3}.
  \label{t0<}
\end{equation}
The momentum (\ref{qx}) decreases when the start time $t_{0}>t_{0<}$ moves away from the left end of the favorable interval. If it is recalled that trajectories with very small momenta are strongly perturbed by the Coulomb field and do not contribute to the process of capture, then the upper end of the window can be estimated from the equation $q_{x}\left(t_{0>},v_{0}=0,\tau_{L}=0\right)$. For $\lambda<1$, one easily finds

\begin{equation}
  \omega t_{0>}\approx-\lambda.
  \label{t0>}
\end{equation}
For the parameters of Fig. \ref{cockroach}, the predictions of Eqs.~ (\ref{t0<}) and (\ref{t0>}), i.e. $\omega t_{0<}\approx-15^{\circ}$ and $\omega t_{0>}\approx-9^{\circ}$, respectively, agree nicely with the results of the Monte-Carlo simulations presented in the lower panel of this figure.

It can be seen that the transverse width of the crescent-like shape in Fig. \ref{cockroach}b, i.e. the maximum transverse velocity allowing for the capture of electrons into the bound states, is more or less the same for start times within the favorable window. To estimate this width, we consider the condition (\ref{conditionest}) for the case when $q_x(t_{0>},v_0,\tau_{L})=0$ and for times in the middle of the laser pulse so that $\tau_L-t_0\approx\tau_L/2$. With account of (\ref{qy}), it immediately takes the form $|v_0|<v_{\max}$ where

\begin{equation}
  v_{\max}=\left(\frac{4}{\tau_{L}}\right)^{1/3}/\left(1-2\frac{F}{\left(2I\right)^{2}}\right).
  \label{vmax}
\end{equation}
Actually, in (\ref{vmax}) we have neglected terms of the order of $\lambda^{2}\ll1$. In Fig. \ref{cockroach}, the transverse velocity does not exceed 0.30 a.u. whereas, for the same parameters, one has from Eq. (\ref{vmax}) $v_{\max}= 0.23$ a.u. Again, the agreement is quite reasonable.

By roughly estimating the capture area in Fig. \ref{cockroach} as $\Sigma^{\ast} \approx\omega\left(t_{0>}-t_{0<}\right)\times2v_{\max}$ and dividing it by the area $\Sigma^{+}$ effective for ionization (see above) we have for the relative yield of neutral excited atoms with respect to singly charged ions

\begin{equation}
  \frac{N^{\ast}}{N^{+}}\propto \frac{\omega}{F^{3/2} \tau_{L}^{2/3}}
  \left(1-2\frac{F}{\left(2I\right)^{2}}\right)^{-1}.
  \label{relyield}
\end{equation}
The scaling with the laser  parameters given by Eq. (\ref{relyield}) was correlated with the results of the Monte-Carlo simulations. According to the latter, when the field strength varies from $F=0.1$ a.u. to  $F=0.07$ a.u. the ratio $N^{\ast}/N^{+}$ rises from $0.18$ to $0.41$, i.e. by a factor of $2.3$, while Eq. (\ref{relyield}) predicts a factor of $1.7$. Agreement is even better with respect to pulse duration. When the pulse duration is doubled from 5 to 10 cycles, then for the field strength of $F=0.1$ a.u. the relative yield drops from $0.18$ to $0.11$, i.e. by a factor of $1.64$, which is practically equal to the factor of $2^{2/3}$ expected from (\ref{relyield}). The general tendency of an increasing percentage of bound electrons for shorter pulses and lower intensities was observed  in the Monte-Carlo simulations of Ref. \cite{Nubbemeyer}.

Experimentally, a strong decrease of the yield of neutral excited atoms with increasing pulse duration was reported in an early paper \cite{Jones93}. When the scaling of Eq. (\ref{relyield}) is applied to analyze
ionization of He atoms with their high ionization potential, the very last factor can be dropped. Then it follows from (\ref{relyield}) that an-order-of-magnitude increase of the laser intensity will reduce the relative yield by a factor of 5 to 6. The relative yield reported in Ref. \cite{Nubbemeyer} decreases over such a range by approximately a factor of 2. The origin of this discrepancy is not presently clear.

\begin{figure}[t]
\begin{center}
\includegraphics[width=0.5\textwidth]{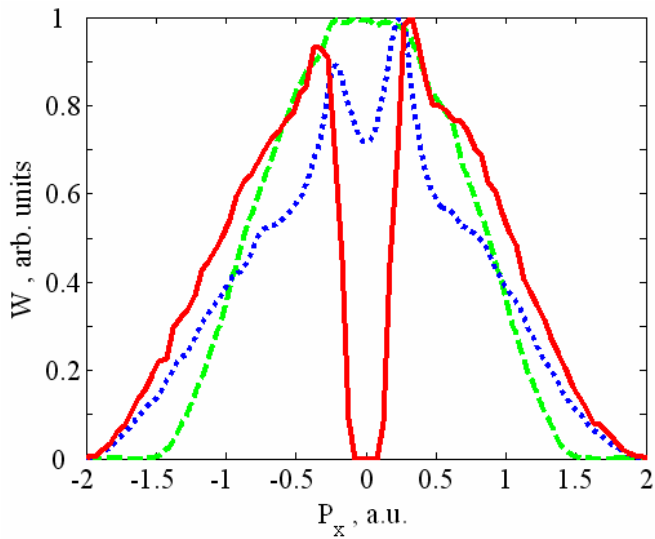}
\end{center}
\caption{The longitudinal momentum distribution evaluated by Monte-Carlo simulations at the end of the laser pulse at time $\tau_{L}$. Dotted blue curve: All launched trajectories, including those with $E(\tau_L)<0$, are accounted for. Solid red curve: The trajectories with negative energies are excluded. Dashed green curve: The result of simulations with the Coulomb field entirely turned off. The parameters are those of Fig. 1.}
\label{sfadistr}
\end{figure}

\begin{figure}[t]
\begin{center}
\includegraphics[width=0.5\textwidth]{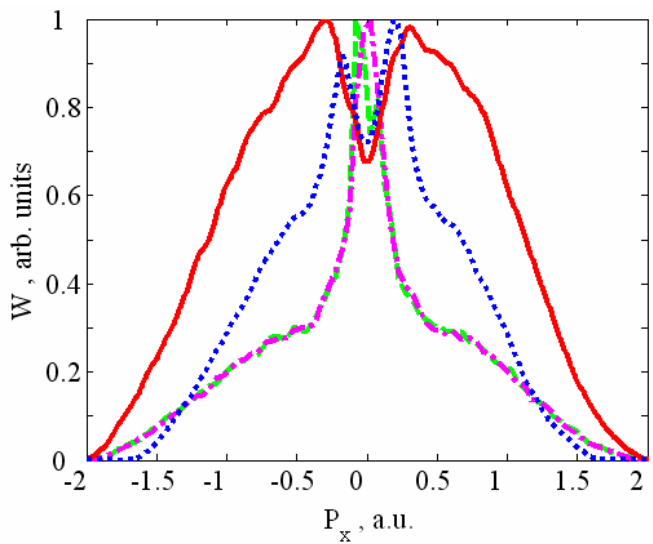}
\end{center}
\caption{The longitudinal momentum distributions, calculated taking into consideration all launched trajectories, at $t=2\tau_{L}$ (dotted blue curve), $t=3 \tau_{L}$ (dashed green curve), $t=4 \tau_{L}$ (dash-dotted magenta curve). The red curve shows the asymptotic distribution, which was evaluated by excluding the trajectories with $E(\tau_L)<0$, as discussed in the text.}
\label{incordistr}
\end{figure}

According to \cite{Nubbemeyer}, the maximum of the $n$-distribution scales with $\sqrt{F}/\omega$. The derivation of this dependence presented there was based on the heuristic assumption that the radial expectation value of the final Rydberg atom, $\left\langle r\right\rangle\propto n^{2}$, is proportional to the amplitude of electron oscillation in the laser field, $A_0=F/\omega^{2}$. Our idea is that the quantum-mechanical mean distance of a captured orbit is determined by the electron distance (\ref{dist}) at the end of the laser pulse, $\left\langle r\right\rangle\approx r\left(t_{0}\right)$. From Monte-Carlo simulations with the parameters of Fig. \ref{enrhist}, we found that the distribution of the distances of trapped electrons (not shown here) has a maximum  around $r_{\max}= 60$ a.u. 
The relation $r_{\max}=\left\langle r\right\rangle=3n^{2}/2$ then gives $n_{\max}\approx6$, which agrees nicely with the results of Fig. \ref{distrn}. The distance (\ref{dist}) indeed becomes proportional to the oscillation amplitude if one factors out the dimensional factor $F/\omega$ from the momentum and converts time into the field phase.

\subsection{The capture into Rydberg states and the distribution of the ionized electrons}

The fact that a substantial part of the tunneled electrons end up in bound states with $E<0$ after the end of the pulse should affect the energy-angular spectrum of direct ionization, particularly its low-energy part. This raises the  question of a possible  relation between the capture into Rydberg states and the dip (minimum) at $p=0$ in the momentum distribution of ionized electrons along the laser polarization, which was for the first time observed in \cite{Mosh03}. The origin of this dip has been the object of intense investigations \cite{Chen02,Dimitriou,Rudenko04,Faisal05,Alnaser06,Mahar06}. In this section we analyze momentum distributions of the ionized electrons, calculated taking into account the capture into bound states.

First, however, we address the photoelectron momentum distribution \textit{at the end of the pulse (not at infinity)} (see Fig. \ref{sfadistr}), calculated with the assumption that all electrons are ionized, regardless of whether the total energy $E(\tau_L)$ [Eq.~(\ref{condition})] is positive or negative \cite{Chen02}. 
Note, that in the presence of the Coulomb field such a distribution is not an observable, since the momentum is not conserved.
However, the momentum distribution at the end of the pulse is instructive for understanding the influence of the Coulomb force on the electron dynamics.
The distribution evaluated without taking the Coulomb field of the atomic residual into account is also shown in the figure. Figure \ref{sfadistr} demonstrates that the Coulomb force severely modifies the central part of the longitudinal momentum distribution. Incidentally, the asymmetry of the dip is due to the small (5 optical cycles) duration of the laser pulse.

\begin{figure}[t]
\begin{center}
\includegraphics[width=0.5\textwidth]{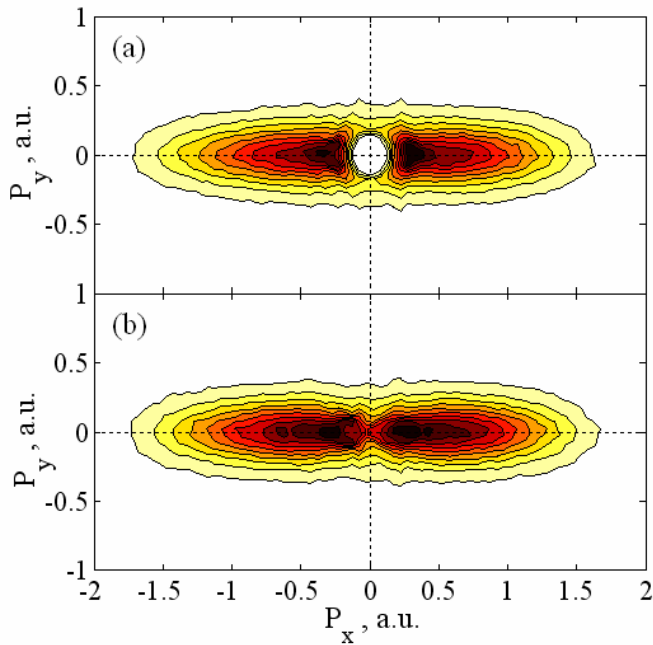}
\end{center}
\caption{(a) The 2D momentum distribution for the parameters of Fig. 1 resulting from Monte-Carlo simulations at the end of the laser pulse, where the trajectories with negative total energy $E=\vecq^2/2-1/r<0$ are discarded. (b) The same distribution in the asymptotic limit after its subsequent evolution in the Coulomb field.}
\label{ourhole}
\end{figure}

Figure \ref{incordistr} illustrates the evolution of the
distribution of $p_{x}$ calculated considering all launched
trajectories, including those with $E(\tau_L)<0$ at the end of the pulse, under the action of the
Coulomb force. One observes that the dip in the longitudinal
distribution is gradually filled in and a narrow maximum develops
in its place. Our simulations show that the dip in this approach
also vanishes with increasing  laser intensity and
pulse duration. We should focus our attention on the fact that the
latter result coincides with a conclusion of Ref.
\cite{Dimitriou}, which, however, concerned the \emph{asymptotic}
distribution of the ionized electrons. It is very important to note that the
\textit{asymptotic} distribution, by which we understand the distribution
to be measured by a detector far away from the ion, has to be calculated \textit{by excluding the trajectories with} $E(\tau_L)<0$. Since the latter are trapped into Ryd\-berg states they will never reach the detector. If nevertheless they are included into the asymptotic distribution, their momenta will not converge to definite values as time goes to infinity or, depending on the numerical accuracy, they will contribute to a spurious accumulation of events with very low momenta. Indeed, the asymptotic
distribution obtained \textit{by excluding the trajectories with
negative energy} (the solid red curve in Fig.~\ref{incordistr}) has a distinct minimum at zero longitudinal momentum.

Let us summarize the results shown in Figs. \ref{sfadistr} and  \ref{incordistr}. A dip in the longitudinal momentum distribution generated by a relatively short laser pulse is already seen at the end of the pulse. A comparison of the two curves calculated with and without the Coulomb field suggests that the Coulomb field is responsible for the formation of this dip \cite{Chen02,Dimitriou}. However, Fig. \ref{incordistr} shows that this dip is washed out if we allow the distribution to continue developing under the action of the Coulomb field after the end of the laser field.  Our definition of the asymptotic distribution discards the contribution of those electrons with negative energy at the end of the pulse. Remarkably, the asymptotic distribution again exhibits a dip. The obvious conclusion is that the mechanism responsible for the formation of the dip at zero momentum is closely related to the one that governs the electron capture into Rydberg states. Both effects are caused by the Coulomb field and occur in adjacent energy intervals.
% But the instability of the double-hump structure with respect to time evolution, demonstrated by Fig. \ref{incordistr}, its smearing out with the increasing of intensity and pulse duration leads to the conclusion that the mechanism underlying the dip formation cannot be traced down to the distortion of the electron trajectory by a Coulomb force.
We notice in passing that the observed structure also survives focal averaging, which is inevitable in real experimental conditions.

Hence, the calculation of the momentum distribution of ionized electrons must take into account the possible population of Rydberg orbits with negative total energy. The two-dimensional momentum distribution, which disregards electrons with $E(t)<0$ at $t=\tau_{L}$, is depicted in Fig. \ref{ourhole}(a). One can see from the figure that the central part of the distribution is completely unpopulated. Simulations for $t>\tau_L$ show that the subsequent motion in the Coulomb field partly fills in this hole, see below.

\begin{figure}[t]
\begin{center}
\includegraphics[width=0.5\textwidth]{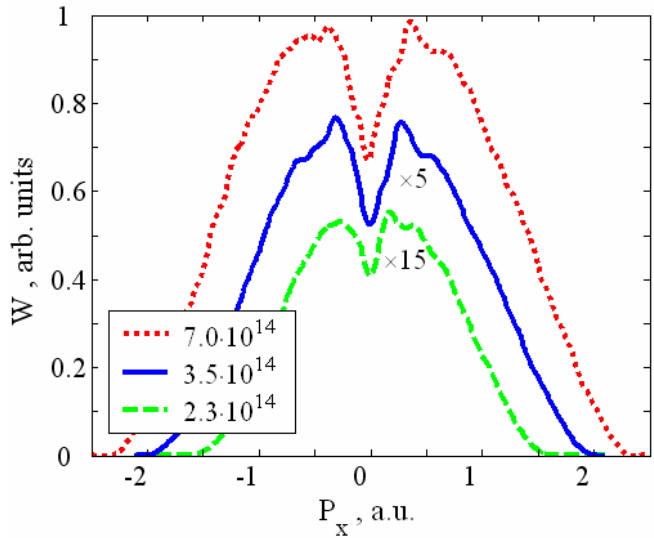}
\end{center}
\caption{Asymptotic momentum distributions along laser polarization for different laser intensities (in W/cm$^{2}$) for the parameters of Fig.1.}
\label{ourint}
\end{figure}

The asymptotic distribution at $t\rightarrow\infty$ can be obtained without calculations up to a large time instant $t_\mathrm{end}$, which guarantees the asymptotic regime for the vast majority of launched trajectories. Indeed, the electron momentum $\vecq(t_0,v_0,\tau_L)$ and its position vector $\vecr(t_0,v_0,\tau_L)$ at the end of the laser pulse uniquely determine the subsequent trajectory in the Coulomb field of the parent ion. Standard formulas of classical mechanics for hyperbolic motion (see, for example \cite{Landau}), allow straightforward analytical evaluation of the electronic asymptotic momentum
\begin{equation}
  \vecp=p\cdot\frac{p\left(\vecA\times\vecM\right)-\vecA}{1+p^2M^2},
    \label{assymp}
\end{equation}
where $\vecM=\vecr\times\vecq$ and $\vecA=\vecq\times\vecM-\vecr/r$ are the conserved angular momentum and Runge-Lenz vector, respectively. The absolute value $p$ of the asymptotic momentum can be determined from the energy conservation law:
\begin{equation}
    \frac{\vecq^{2}}{2}-\frac{1}{r}=\frac{p^2}{2}.
    \label{holeexp}
\end{equation}

It should be noticed that Eq.~(\ref{holeexp}) can also be used to explain the origin of the unpopulated area in Fig. \ref{ourhole}(a): In fact, at the end of the laser pulse electrons can be found at various finite distances from the ion. The positivity of the total energy imposes a lower limit  on the value of the electron momentum at that instant of time. In contrast, the electron does not feel the Coulomb field at $t\rightarrow\infty$, and its momentum can be arbitrarily small.

The distribution evaluated by this approach is shown in Fig. \ref{ourhole}(b). Together with Fig. \ref{ourint} it demonstrates that the dip is not smeared out by the evolution in the Coulomb field after the end of the laser pulse and, moreover, its shape is not very sensitive to the laser intensity. The same holds for the behavior of the distribution with increasing pulse duration.

Thus, depletion of the low-energy part of the ionization spectrum due to the capture into Rydberg states is a possible mechanism for the formation of a dip in the momentum distribution along the polarization direction.

\section{Conclusions}

Based on the two-step semiclassical model, we have investigated the electron motion after tunneling in a strong laser field considering the interaction with the Coulomb field of the parent ion.  The focus of attention is on the mechanism governing the yield of neutral excited atoms from a gaseous target irradiated by a short laser pulse. A simple physical picture emerges from Monte-Carlo simulations with classical trajectories after tunneling. With account of the Coulomb field, not all tunneled electrons are actually ionized since some of them are captured back into bound atomic states after the end of the laser pulse. 
Obviously, an electron remains bound if its total energy at the end of the laser pulse is negative. Furthermore, we show that for an electron to be captured it (i) has to have moderate drift momentum  at the time of tunneling and (ii) has to avoid strong interaction (``hard'' collision) with the ion.  Such trajectories do exist when electrons are released within a narrow time window before the extrema of the oscillating laser field having a not too large transverse velocity $\vecv_0$. 
Understanding these features has allowed us to derive the scaling of the ratio of neutral excited atoms and singly charged ions with the laser--atom parameters in analytic form. Its prediction,
that the percentage of neutrals increases with decreasing intensity or pulse duration, fits into the general physical picture: In both cases, the electron drifts away from the ion by a smaller distance during the laser pulse and experiences a stronger Coulomb attraction. This distance is smaller at  lower intensity because of a reduced drift momentum imparted by the laser field and, in the other case of a shorter pulse, because of a smaller travel time.  In addition, it has been demonstrated that the channel of capture, when accounted for in semiclassical calculations, has a pronounced effect on the momentum distribution of electrons with small positive energy. In particular, it is correlated with the formation of a ``dip'' at zero momentum in the distribution of the longitudinal momentum.

\section*{Acknowledgments}
The authors would like to thank U.~Eichmann, M. V.~Fedorov, and A. M.~Popov for
stimulating and useful discussions. This work was in part supported by the
Deutsche Forschungsgemeinschaft and by the Russian Foundation for Basic Research.

%%%%%%%%%%%%%%%%Bibliography environment%%%%%%%%%%%%%%%

\end{document}